\documentclass[preprint,superscriptaddress]{revtex4-1}

\usepackage{amsmath}    
\usepackage{graphicx}   
\usepackage{verbatim}   
\usepackage{color}      
\usepackage{subfigure}  
\usepackage{hyperref}   
\usepackage{dcolumn}
\usepackage{bm}
\usepackage{multirow}

\def\dblone{\hbox{$1\hskip -1.2pt\vrule depth 0pt height 1.6ex width 0.7pt\vrule depth 0pt height 0.3pt width 0.12em$}}

\begin{document}
\title{Faddeev Random Phase Approximation for Molecules}
\date{\today}
\author{Matthias Degroote}
\author{Dimitri Van Neck}
\affiliation{Center for Molecular Modeling, Technologiepark 903, B-9052 Zwijnaarde, Belgium}
\affiliation{Members of the Ghent-Brussels Quantum Chemistry and Molecular Modeling Alliance}
\author{Carlo Barbieri}
\affiliation{Department of Physics, Faculty of Engineering and Physical Sciences, University of Surrey, Guildford GU2 7XH, United Kingdom}

\begin{abstract}
The Faddeev Random Phase Approximation is a Green's function technique that makes use of Faddeev-equations to couple the motion of a single electron to the two-particle--one-hole and two-hole--one-particle excitations. This method goes beyond the frequently used third-order Algebraic Diagrammatic Construction method: all diagrams involving the exchange of phonons in the particle-hole and particle-particle channel are retained, but the phonons are described at the level of the Random Phase Approximation. This paper presents the first results for diatomic molecules at equilibrium geometry. The behavior of the method in the dissociation limit is also investigated.
\end{abstract}

\maketitle
\section{Introduction}
The study of electronic systems by means of first-principle calculations has taken a high rise thanks to modern computer technology~\cite{White1992,Chan2008,Bartlett1981,Ceperley1996,Acioli1997}. The Green's function formalism~\cite{Pines1962,Fetter1971,Dickhoff2008} is one of these first-principles methods that has been succesfully applied in quantum chemistry~\cite{Linderberg1973,Cederbaum1977,Vonniessen1984,Ortiz1997}. The correlations in a many-body system are described in terms of  an electron self-energy that acts as an energy-dependent potential describing the motion of a single electron in the many-electron system.

A particular third-order approximation scheme to the self-energy can be obtained using the Algebraic Diagrammatic Construction (ADC(3))~\cite{Schirmer1983} method as developed by Schirmer and coworkers. This method has proven to be very successful in predicting one-electron properties in molecules~\cite{Deleuze1999} as measured e.g. in electron momentum spectroscopy. Although the equations were derived in a purely algebraic manner, they can be shown to be equivalent to resumming all particle-hole (ph) and particle-particle (pp) interactions between two-particle--one-hole (2p1h) and two-hole--one-particle (2h1p) states up to the Tamm-Dancoff approximation (TDA)~\cite{Fetter1971} level. This is diagrammatically equivalent to considering phonons (excitations in the ph and pp channel) at the TDA level, and then allowing the exchange of these phonons in all possible ways between the tree propagators describing the 2p1h/2h1p states.

The TDA allows no ground-state correlations in the construction of the phonons. An improvement in this respect is the Random Phase Approximation (RPA)~\cite{Ring}. Calculations for the electron gas show that these improvements lead to a correct prediction of the plasmon pole, whereas the TDA completely fails to describe the plasmon spectrum. It is therefore of interest to formulate an analogous theory to ADC(3) that resums the ph and pp interactions up to RPA level.

Going beyond the TDA level has proven to be very difficult~\cite{Rijsdijk1996}, even though it is known that the RPA should be better to describe collective behavior, at least for nuclear systems~\cite{Ring}. The Faddeev Random Phase Approximation (FRPA)~\cite{Barbieri2001} solves this problem by using the Faddeev technique to include RPA-phonons in the self-energy. This method has succesfully been applied to both nuclei~\cite{Barbieri2002,Barbieri2009} and atoms~\cite{Barbieri2007}. It is the aim of the present paper to study the application of this technique to simple molecular systems.

In the second section of this work we give a short overview of the working equations for the FRPA method. In section~\ref{sec:results} we present the numerical results for a set of diatomic molecules. A summary is provided in section~\ref{sec:conclusion}.

\section{Theory}
\subsection{Single-particle Green's Function}
The single-particle motion in an N-body system is described by the single-particle propagator
\begin{equation}
G_{\alpha,\beta}\left(t,t'\right) = -\frac{i}{\hbar}\left<\Psi_0^N\left| \mathcal{T}\left[a_{\alpha}(t)a_{\beta}^{\dagger}(t')\right]\right|\Psi_0^N\right>
\end{equation}
where $\mathcal{T}[...]$ represents the time-ordering operator, $\Psi_0^N$ is the exact ground state and $a_{\alpha}(t)$ and $a_{\alpha}^{\dagger}(t)$ are the addition and removal operators in the Heisenberg representation for an electron in a single-particle state $\alpha$. For practical calculations it is more convenient to use the Lehmann representation of the Green's function
\begin{eqnarray}
G_{\alpha,\beta}\left(E\right) &=& \sum_{m>F} \frac{\left<\Psi_0^N\left| a_{\alpha}\right|\Psi_m^{N+1}\right>\left<\Psi_m^{N+1}\left|a_{\beta}^{\dagger}\right|\Psi_0^N\right>}{E-(E_m^{N+1}-E_0^N) +i\eta} + \sum_{m<F} \frac{\left<\Psi_0^N\left| a^{\dagger}_{\alpha}\right|\Psi_m^{N-1}\right>\left<\Psi_m^{N-1}\left|a_{\beta}\right|\Psi_0^N\right>}{E-(E_0^{N}-          E_{m}^{N-1}) -i\eta}\nonumber\\
&=& \sum_{m>F} \frac{f_{\alpha,m}f_{\beta,m}^{*}}{E-\omega_m + i \eta} + \sum_{m<F} \frac{f_{\alpha,m}f_{\beta,m}^{*}}{E-\omega_m -i \eta}, \label{eqn:gleh}
\end{eqnarray}
where the $\Psi_m^{N\pm1}$ represent exact eigenstates of the Hamiltonian with energy $E_m^{N\pm1}$. This transition to the energy domain transforms the Dyson equation from an integral equation into the algebraic relation
\begin{equation}
G_{\alpha,\beta}\left(E\right) = G^{(0)}_{\alpha,\beta}\left(E\right) + \sum_{\gamma,\delta} G^{(0)}_{\alpha,\gamma}\left(E\right)\Sigma^{*}_{\gamma,\delta}\left(E\right)G_{\delta,\beta}\left(E\right).
\label{eqn:g}
\end{equation}
In this equation the exact Green's function $G$ is expressed in terms of the non-interacting $G^{(0)}$ and the irreducible self-energy $\Sigma^{*}(E)$. Approximation schemes for the single-particle Green's function boil down to finding an appropriate perturbation expansion for the irreducible self-energy.

\begin{figure}
   \begin{center}
      \includegraphics{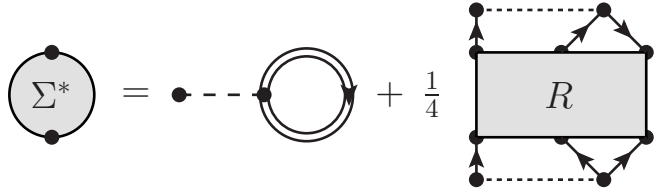}
      \caption{The Feynman-diagram for the irreducible self-energy $\Sigma^{*}$ in equation~(\ref{eqn:sigma}) within the FRPA. The first diagram represents the HF-like static self-energy.}
      \label{fig:jaxo1}
   \end{center}
\end{figure}

In our approach, we want to couple the single-particle states with 2p1h and 2h1p states. According to Refs. \cite{Ethofer1969, Winter1972}, the connection between the irreducible self-energy $\Sigma^*$ and the six-point response function $R$ can be written as
\begin{equation}
\Sigma^{*} _{\alpha,\beta}\left(E\right) = \Sigma^{HF}_{\alpha,\beta} + \frac{1}{4} \sum_{\lambda,\mu,\nu} \sum_{\epsilon,\theta,\sigma} V_{\alpha\nu,\lambda\mu}R_{\lambda\mu\nu,\epsilon\theta\sigma}\left(E\right)V_{\epsilon\theta,\beta\sigma}
\label{eqn:exact}
\end{equation}
where $V$ is the anti-symmetrized two-particle interaction and $\Sigma^{HF}$ is the static self-energy as depicted in Figure~\ref{fig:jaxo1}. We now replace the exact single-energy six-point response function $R(E)$ by an approximate propagator that has indices that are restricted to the 2p1h space ($R^{2h1p}$) or 2h1p space ($R^{2h1p}$), and that is exact up to third order:
\begin{equation}
\Sigma^{*}_{\alpha,\beta}\left(E\right) = \Sigma^{HF}_{\alpha,\beta} + \frac{1}{4} \sum_{\lambda,\mu,\nu} \sum_{\epsilon,\theta,\sigma} U_{\alpha\nu,\lambda\mu}R_{\lambda\mu\nu,\epsilon\theta\sigma}\left(E\right)U_{\epsilon\theta,\beta\sigma}.
\label{eqn:sigma}
\end{equation}
The two-particle interaction $V$ in Eq.~(\ref{eqn:exact}) has been replaced by a second order expansion
\begin{equation}
U_{\alpha\beta,\gamma\delta} = \sum_{\lambda,\mu}\left(\dblone_{\alpha\beta,\lambda\mu} + \Delta U_{\alpha\beta,\lambda\mu}\right)V_{\lambda\mu,\gamma\delta}.
\label{eqn:U}
\end{equation}
This $\Delta U$ is needed to guarantee full summation up to third order perturbation theory and was chosen to be the same as the vertex correction used in the ADC(3)~\cite{Schirmer1983}.

\subsection{pp/ph RPA interaction}
\begin{figure}
   \begin{center}
      \subfigure[]{
      \includegraphics{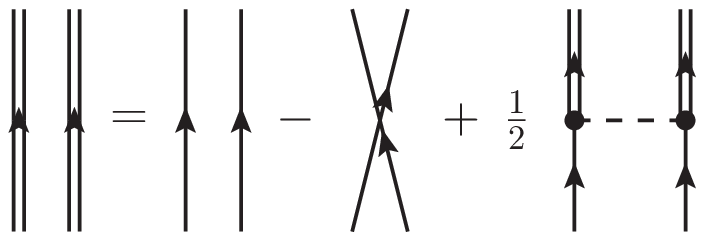}
      \label{fig:ppRPA}
      }
      \subfigure[]{
      \includegraphics{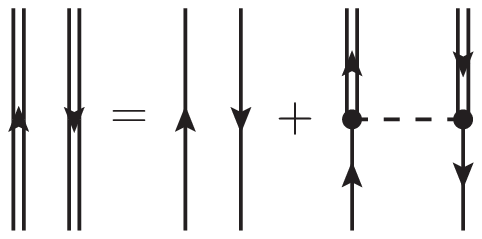}
      \label{fig:phRPA}
      }
   \end{center}
   \caption{The diagrammatical representation of the pp-RPA equation~\subref{fig:ppRPA} and the ph-RPA equation~\subref{fig:phRPA} where the single lines represent non-interacting and the double lines interacting propagators.}
   \label{fig:RPA}
\end{figure}

The two-particle propagator is defined by
\begin{eqnarray}
G^{pp}_{\alpha\beta,\gamma\delta}\left(E\right) &=& \sum_m \frac{\left<\Psi_0^{N}\left|a_{\beta}a_{\alpha}\right|\Psi_m^{N+2}\right>\left<\Psi_m^{N+2}\left|a_{\gamma}^{\dagger}a_{\delta}^{\dagger}\right|\Psi_0^N\right>}{E-(E^{N+2}_m - E^{N}_0)+i\eta} \nonumber\\
&&\quad- \sum_n \frac{ \left<\Psi_0^{N}\left|a_{\gamma}^{\dagger}a_{\delta}^{\dagger}\right|\Psi_n^{N-2}\right>\left<\Psi_n^{N-2}\left|a_{\beta}a_{\alpha}\right|\Psi_0^N\right>}{ E-(E^{N}_0 - E^{N-2}_n)-i\eta} \label{eqn:pplong} \\
&=&  \sum_m \frac{\mathcal{X}_{\alpha\beta,m}^{pp}\mathcal{X}_{\gamma\delta,m}^{pp \dagger}}{E-\epsilon^{pp+}_m+i\eta} - \sum_n \frac{ \mathcal{Y}^{pp}_{\gamma\delta,n}\mathcal{Y}^{pp\dagger}_{\alpha\beta,n}}{ E-\epsilon^{pp-}_n-i\eta},
\end{eqnarray}
where the $\mathcal{X}^{pp}$, $\mathcal{Y}^{pp}$ and $\epsilon^{pp}$ are shorthand notations for the overlap amplitudes and energy differences in Eq.~(\ref{eqn:pplong}). A relevant approximation for this object is obtained by solving the RPA equations~\cite{Ring}
\begin{eqnarray}
G^{pp}_{\alpha\beta,\gamma\delta}\left(E\right) &=& G^{pp(0)}_{\alpha\beta,\gamma\delta}\left(E\right) + \frac{1}{2}\sum_{\lambda\mu}G^{pp(0)}_{\alpha\beta,\alpha\beta}\left(E\right)V_{\alpha\beta,\lambda\mu}G^{pp}_{\lambda\mu,\gamma\delta}\left(E\right)\\\label{eqn:RPA1}
&=& G^{pp(0)}_{\alpha\beta,\gamma\delta}\left(E\right) + G^{pp(0)}_{\alpha\beta,\alpha\beta}\left(E\right)\Gamma^{pp}_{\alpha\beta,\gamma\delta}\left(E\right)G^{pp(0)}_{\gamma\delta,\gamma\delta}\left(E\right),\label{eqn:ppint}
\end{eqnarray}
as indicated diagrammatically in Figure~\ref{fig:ppRPA}. Equation~(\ref{eqn:ppint}) defines the effective pp interaction $\Gamma^{pp}$, which includes dynamical screening and will be used later as a building block for the 2p1h and 2h1p interaction. This simple form of the Bethe-Salpeter-like equation for the pp propagator in function of a screened interaction $\Gamma^{pp}$ is possible because the non-interacting pp propagator is diagonal in the HF basis.

The same procedure can be followed for the particle-hole (ph) polarization propagator (see Figure~\ref{fig:phRPA}), defined as
\begin{eqnarray}
&=& \sum_m \frac{\left<\Psi_0^N\left| a_{\beta}^{\dagger}a_{\alpha} \right|\Psi_m^{N}\right>\left<\Psi_m^N\left|a_{\gamma}^{\dagger}a_{\delta}\right|\Psi_0^N\right>}{E-(E_m^N-E_0^N) + i\eta} \nonumber\\
&&\quad- \sum_n \frac{\left<\Psi_0^N\left| a_{\gamma} ^{\dagger}a_{\delta}\right|\Psi_n^{N}\right>\left<\Psi_n^N\left|a_{\beta}^{\dagger}a_{\alpha}\right|\Psi_0^N\right>}{E-(E_0^N - E_n^N) - i\eta}\\
&=& \sum_m \frac{\mathcal{X}^{ph}_{\alpha\beta,m}\mathcal{X}^{ph\dagger}_{\gamma\delta,m}}{E-\epsilon^{ph+}_m + i\eta} - \sum_n \frac{\mathcal{Y}^{ph\dagger}_{\alpha\beta,n}\mathcal{Y}^{ph}_{\gamma\delta,n}}{E-\epsilon^{ph-}_n -i\eta}.
\end{eqnarray}

The corresponding Bethe-Salpeter-like equation in the RPA reads as
\begin{eqnarray}
\Pi^{ph}_{\alpha\beta,\gamma\delta}\left(E\right) &=& \Pi^{ph(0)}_{\alpha\beta,\gamma\delta}\left(E\right) + \sum_{\lambda,\mu} \Pi^{ph(0)}_{\alpha\beta,\gamma\delta}\left(E\right)V_{\alpha\mu,\beta\lambda}\Pi^{ph}_{\lambda\mu,\gamma\delta}\left(E\right)\\\label{eqn:RPA2}
&=& \Pi^{ph(0)}_{\alpha\beta,\gamma\delta}\left(E\right) + \Pi^{ph(0)}_{\alpha\beta,\alpha\beta}\left(E\right)\Gamma^{ph}_{\alpha\beta,\gamma\delta}\left(E\right)\Pi^{ph(0)}_{\gamma\delta,\gamma\delta}\left(E\right), \label{eqn:phint}
\end{eqnarray}
and defines the effective ph interaction $\Gamma^{ph}$.

The actual calculation of the amplitudes and poles of the pp propagator and ph polarization propagator can be done by solving the generalized eigenvalue problems~\cite{Ring}
\begin{equation}
\label{eqn:RPApp}
\left(\begin{array}{cc} A & B \\ B^{\dagger} & C\end{array}\right) \left(\begin{array}{cc} \mathcal{X}^{pp+} & \mathcal{Y}^{pp-}  \\ \mathcal{Y}^{pp+} & \mathcal{X}^{pp-} \end{array}\right) = \left(\begin{array}{cc} \dblone & 0 \\ 0 & -\dblone\end{array}\right)\left(\begin{array}{cc}\mathcal{X}^{pp+} & \mathcal{Y}^{pp-}  \\ \mathcal{Y}^{pp+} & \mathcal{X}^{pp-}\end{array}\right)\left(\begin{array}{cc}\epsilon^{pp+} & 0 \\ 0 & \epsilon^{pp-}\end{array}\right)
\end{equation}
where
\begin{align}
A_{\alpha\beta,\gamma\delta} = & \left(\delta_{\alpha\gamma}\delta_{\beta\delta}-\delta_{\alpha\delta}\delta_{\beta\gamma}\right)\left(\epsilon_{\alpha}+\epsilon_{\beta}\right) + \frac{1}{2}V_{\alpha\beta,\gamma\delta} & \alpha,\beta,\gamma,\delta > F\\
B_{\alpha\beta,\gamma\delta} =& V_{\alpha\beta,\delta\gamma} & \alpha,\beta > F; \gamma,\delta < F\\
C_{\alpha\beta,\gamma\delta} = & \left(\delta_{\alpha\gamma}\delta_{\beta\delta}-\delta_{\alpha\delta}\delta{\beta\gamma}\right)\left(\epsilon_{\alpha}+\epsilon_{\beta}\right) - \frac{1}{2}V_{\alpha\beta,\gamma\delta} & \alpha,\beta,\gamma,\delta < F.
\end{align}
Here the $\epsilon_{\alpha}$ represent Hartree-Fock single-particle energies with the Fermi level $F$ separating the occupied and unoccupied HF levels. The equations for the ph polarization propagator are again very similar:
\begin{equation}
\label{eqn:RPAph}
\left(\begin{array}{cc} D & E \\ E^{\dagger} & D^{\dagger}\end{array}\right) \left(\begin{array}{cc} \mathcal{X}^{ph+} & \mathcal{Y}^{ph-} \\ \mathcal{Y}^{ph+} & \mathcal{X}^{ph-} \end{array}\right) = \left(\begin{array}{cc} \dblone & 0 \\ 0 & -\dblone\end{array}\right)\left(\begin{array}{cc}\mathcal{X}^{ph+} & \mathcal{Y}^{ph-}  \\ \mathcal{Y}^{ph+} & \mathcal{X}^{ph-}\end{array}\right)\left(\begin{array}{cc}\epsilon^{ph+} 0 \\ 0 & \epsilon^{ph-} \end{array}\right)
\end{equation}
where
\begin{align}
D_{\alpha\beta,\gamma\delta} =& \delta_{\alpha\gamma}\delta_{\beta\delta}\left(\epsilon_{\alpha}-\epsilon_{\beta}\right) + V_{\alpha\delta,\beta\gamma}& \alpha,\gamma > F; \beta, \delta < F\\
E_{\alpha\beta,\gamma\delta} =& V_{\alpha\gamma,\beta\delta}& \alpha,\gamma > F; \beta,\delta < F.
\end{align}
\subsection{Faddeev equations}
The diagrammatic content of $R$ cannot be cast into the form of a Bethe-Salpeter equation without double counting of some classes of diagrams, in contrast to the more complicated 4-times propagator (see Ref.~\cite{Barbieri2001}). That is why the Faddeev technique~\cite{Faddeev} must be used to split this object into three parts. The analysis will be done for $R^{2p1h}$ (the derivation of $R^{2h1p}$ is found to be completely analogous, but with an interchange of particle and hole lines). The decomposition of $R^{2p1h}$ into three Faddeev components $R^{(i)}$ reads
\begin{equation}
\label{eqn:R2p1h}
R^{2p1h}_{\alpha\beta\gamma,\lambda\mu\nu}\left(E\right) = G^{(0)>}_{\alpha\beta\gamma,\lambda\mu\nu}\left(E\right) - G^{(0)>}_{\alpha\beta\gamma,\mu\lambda\nu}\left(E\right)+ \sum_{i=1,2,3}R^{ (i) }_{\alpha\beta\gamma,\lambda\mu\nu}\left(E\right),
\end{equation}
where $G^{(0)>}$ is the part of the non-interacting 2p1h propagator with positive energy
\begin{equation}
G^{(0)>}_{\alpha\beta\gamma,\lambda\mu\nu}\left(E\right) = \frac{\delta_{\alpha\lambda}\delta_{\beta\mu}\delta_{\gamma\nu}}{E-\left(\epsilon_{\alpha}+\epsilon_{\beta}-\epsilon_{\gamma}\right)+i\eta}.
\end{equation}
Together with its exchange counterpart, they form the free 2p1h propagator
\begin{equation}
R^{free}_{\alpha\beta\gamma,\lambda\mu\nu}\left(E\right) = G^{(0)>}_{\alpha\beta\gamma,\lambda\mu\nu}\left(E\right) - G^{(0)>}_{\alpha\beta\gamma,\mu\lambda\nu}\left(E\right)
\end{equation}

\begin{figure}
   \begin{center}
      \includegraphics[width=\textwidth]{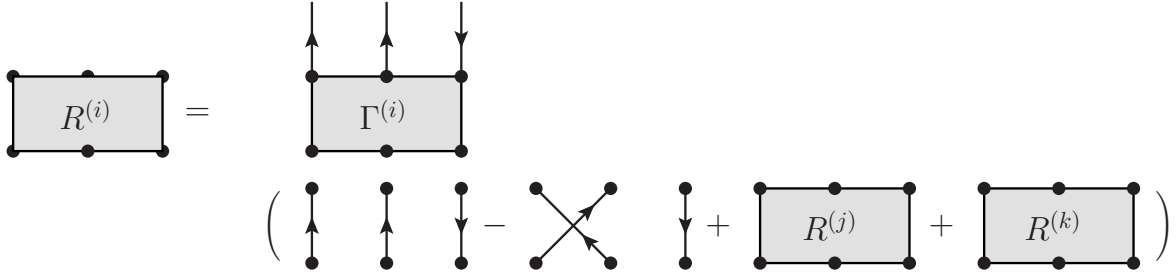}
      \caption{Diagrammatic representation of equation~(\ref{eqn:FRPAsc}).}
      \label{fig:Ri}
   \end{center}
\end{figure}

The relation between the different components $R^{(i)}$ can be derived from the diagrammatic content of Figure~\ref{fig:Ri}. The superscripts (i), (j) and (k) are cyclical permutations of 1, 2 and 3 and correspond to the numbering of the fermion lines from left to right. In our notation lines 1 and 2 are the particles and line 3 is the hole. Each propagator $R^{(i)}$ ends with lines $j$ and $k$ interacting through the adequate RPA interaction vertex, while all possible prior propagation is included in $R^{(j)}$, $R^{(k)}$ and the non-interacting propagators. $\Gamma^{(i)}$ is the extension to 2p1h space of $\Gamma^{pp}$ and $\Gamma^{ph}$ by adding a Kronecker delta for the third fermion line.
The corresponding Bethe-Salpeter equations for the $R^{(i)}$
\begin{eqnarray}
R^{(i)}_{\alpha\beta\gamma,\lambda\mu\nu}\left(E\right) &=& \sum_{\zeta\eta,\theta} \left[G^{(0)>}\Gamma^{(i)}\right]_{\alpha\beta\gamma,\zeta\eta\theta}\left(E\right)\left(G^{(0)>}_{\zeta\eta\theta,\lambda\mu\nu}\left(E\right) - G^{(0)>}_{\zeta\eta\theta,\mu\lambda\nu}\left(E\right)\right.\nonumber\\
&&\quad\left.+ R^{(j)}_{\zeta\eta\theta,\lambda\mu\nu}\left(E\right) + R^{(k)}_{\zeta\eta\theta,\lambda\mu\nu}\left(E\right)\right)
\label{eqn:FRPAsc}
\end{eqnarray}
form a closed self-consistent system.

The Lehmann representation 
\begin{equation}
\label{eqn:R}
R^{(i)}_{\alpha\beta\gamma,\lambda\mu\nu} = \sum_{m} \frac{\mathcal{X}^{(i)}_{\alpha\beta\gamma,m}\mathcal{X}_{\lambda\mu\nu,m}}{E-\epsilon^{Fd}_m + i\eta} - R^{(i)free}_{\alpha\beta\gamma,\lambda\mu\nu}
\end{equation}
can be derived from the Lehman representation for the full $R$ (see Ref.~\cite{Barbieri2001}). The sum of the $R^{(i)free}$ makes sure that the non-interacting poles appearing in the first term of Eq.~(\ref{eqn:R2p1h}) are precisely cancelled. The spectroscopic amplitude can be recovered by summing over the three Faddeev components
\begin{equation}
\label{eqn:FRPAsum}
\mathcal{X}_{\alpha\beta\gamma,m} = \sum_{i=1,2,3} \mathcal{X}^{(i)}_{\alpha\beta\gamma,m}.
\end{equation}
By multiplying equation~(\ref{eqn:R}) with $(E-\epsilon^{Fd}_m)$ and taking the limit for $E\rightarrow \epsilon^{Fd}_m$, the problem is reduced to a non-linear eigenvalue problem for the spectroscopic amplitudes and the poles. The non-interacting poles do not coincide with the Faddeev-poles, so the $R^{free}$ is guaranteed to disappear when taking the limit:
\begin{equation}
\label{eqn:eigval}
\mathcal{X}^{(i)}_{\alpha\beta\gamma,m} = \sum_{\zeta<\eta,\theta}\left(G^{(0)>}\Gamma^{(i)}\right)_{\alpha\beta\gamma,\eta\zeta\theta}\left(\epsilon_m^{Fd}\right)\left(\mathcal{X}^{(j)}_{\eta\zeta\theta,m} + \mathcal{X}^{(k)}_{\eta\zeta\theta,m}\right).
\end{equation}
The explicit treatment of this equation for $i=3$ (i.e. the pp channel) is given in appendix~\ref{app:frpa}, and is easily extended to the two other channels. When substituted in equation~(\ref{eqn:eigval}), we arrive at
\begin{equation}
\mathcal{X}^{\left(i\right)} = \left( U^{\left(i\right)}\frac{1}{\epsilon_m^{Fd} - D^{\left(i\right)}}T^{\left(i\right)\dagger} +H^{\left(i\right)}H^{\left(i\right)\dagger}\right)\left(\mathcal{X}^{\left(j\right)} + \mathcal{X}^{\left(k\right)}\right).
\end{equation}
The vectors $U^{\left(i\right)}$, $D^{\left(i\right)}$, $T^{\left(i\right)}$ and $H^{\left(i\right)}$ are all diagonal in the freely propagating line and can be written in terms of the pp- and ph-amplitudes and energies. Their explicit form can be found in Ref.~\cite{Barbieri2001}. By introducing a vector containing these three components,
\begin{equation}
X = \left(\begin{array}{c} \mathcal{X}^{\left(1\right)}\\ \mathcal{X}^{\left(2\right)}\\ \mathcal{X}^{\left(3\right)}\end{array}\right),
\end{equation}
this non-linear equation in the Faddeev-energies and amplitudes can be written in the form
\begin{equation}
X = \left(U\frac{1}{\epsilon^{Fd} - D}T^{\dagger} + HH^{\dagger}\right)MX
\end{equation}
where the matrix M,
\begin{equation}
M = \left(\begin{array}{ccc} 0 & \dblone & \dblone \\ \dblone & 0 & \dblone\\ \dblone & \dblone & 0 \end{array}\right),
\end{equation}
takes care of the coupling between the different channels. After some matrix algebra, this can be converted into a linear non-hermitian eigenvalue problem
\begin{equation}
\label{eqn:FRPAham}
\epsilon^{Fd} X = \left(\dblone - HH^{\dagger}M\right)^{-1}U\left[T^{\dagger}M + D U^{-1}\left(\dblone - HH^{\dagger}M\right)\right]X.
\end{equation}\\
The matrix dimension of the eigenvalue problem is three times the size of the 2p1h-basis. Two thirds of the solutions are spurious and can be projected out, so the actual matrix dimension reduces to the size of a single 2p1h basis.

\subsection{Handling spurious solutions}
The use of Faddeev-equations inherently introduces spurious solutions~\cite{Adhikari1979, Evans1981, Navratil1999}. The solutions for which the sum in Eq.~(\ref{eqn:FRPAsum}) is zero, have no physical meaning and have to be discarded. At the same time the vectors themselves will have to be anti-symmetric under exchange of the two particle or hole lines. By projecting the Hamiltonian matrix~(\ref{eqn:FRPAham}) onto the vector that has the right symmetry properties, and is non-vanishing when summed, the matrix dimension is reduced by a factor of $3$. This vector space is spanned by the vector
\begin{equation}
\frac{1}{\sqrt{6}}\left(\begin{array}{c}\dblone-\dblone_{ex}\\\dblone-\dblone_{ex}\\\dblone-\dblone_{ex}\end{array}\right),
\end{equation}
where $\left(\dblone_{ex}\right)_{\alpha\beta\gamma,\lambda\mu\nu}=\delta_{\alpha\mu}\delta_{\beta\lambda}\delta_{\gamma\nu}$.
The dimension of the matrix is now the same as in the standard ADC(3) matrix problem~\cite{Schirmer1983}. It can be verified that by using Tamm-Dancoff (TDA) interactions and after performing this projection, one regains the ADC(3) equations (see Appendix~\ref{app:adc}).
\subsection{Single particle propagator and ground-state properties}
The calculation of the FRPA single-particle propagator is now done by diagonalization of the symmetric matrix
\begin{equation}
\label{eqn:FRPAmatrix}
\bordermatrix{& \text{p/h} & \text{2p1h} & \text{2h1p} \cr
\text{p/h} & \epsilon & \tilde{U} & \tilde{U} \cr
\text{2p1h} & \tilde{U}^{\dagger} & \epsilon^{Fd} & 0 \cr
\text{2h1p} & \tilde{U}^{\dagger} & 0 & \epsilon^{Fd} }
\end{equation}

where the $\epsilon^{Fd}$ matrices are diagonal and contain the 2p1h and 2h1p Faddeev energies. The tilde indicates that the coupling matrix elements are written in the basis that diagonalizes the Faddeev matrices:
\begin{equation}
\tilde{U}_{\alpha,m} = \sum_{\lambda,\mu,\nu} U_{\alpha\nu,\lambda\mu}\mathcal{X}_{\lambda\mu\nu,m}.
\end{equation}
Note that in standard ADC(3) it is possible to write the equivalent of matrix~(\ref{eqn:FRPAmatrix}) using (\ref{eqn:adc3}) and (\ref{eqn:U}) as sub-blocks without a separate diagonalization in 2p1h and 2h1p space. This is not the case in the FRPA formalism as due to the nonhermiticity of the right hand side of Eq.~\ref{eqn:FRPAham}. Thus, one should first diagonalize the 2p1h and 2h1p sub-blocks (that is, solve the Faddeev equations) and then write the matrix~(\ref{eqn:FRPAmatrix}) in the new basis obtained. Performing the double diagonalization procedure therefore involves a doubling of the computer time with respect to the usual ADC(3) approach. In practical calculations, however, this is not the case since the dimension of matrix~(\ref{eqn:FRPAmatrix}) can be reduced drastically by employing Arnoldi techniques in the 2p1h and 2h1p diagonalizations. This approach has been applied previously~\cite{Barbieri2009} and it was found that a limited number of Arnoldi vectors guarantee correct converged results for total energies and ionizations potentials. In this paper, however, we did not resort to the Arnoldi algorithm and all results are obtained with full diagonalizations.

The diagonalization of (\ref{eqn:FRPAmatrix}) results in energies $\omega_n$ and residues $f_{\alpha,n}$ (see Eq.~(\ref{eqn:gleh})), defining a new single-particle Green's function. By summing over the the solutions below the Fermi level, the density matrix
\begin{equation}
n_{\alpha,\beta} = \sum_{n<F} f_{\alpha,n} f_{\beta,n}^{*}
\end{equation}
and the corresponding ground-state energy
\begin{equation}
E_0^N = \frac{1}{2}\left(\sum_{\alpha,\beta} \left< \alpha \left| T \right| \beta \right>n_{\alpha\beta} + \sum_{\alpha}\sum_{n<F}\omega_n f_{\alpha,n} f_{\alpha,n}^{*}\right)
\end{equation}
can be obtained.

In principle full self-consistency could be achieved by iteratively recalculating the phonons on the basis of the new propagator and applying the Faddeev procedure. This is however computationally too demanding. We do improve the self-consistency of the solution by updating the Hartree-Fock-like static self-energy diagram. Instead of the diagonal matrix of single-particle energies, the Hartree-Fock self-energy calculated with the new density matrix $n_{\alpha,\beta}$ has to be included in the diagonalization. Note that, both in FRPA and ADC(3), this partially self-consistent treatment is needed to include all static self-energy diagrams up to third order.
\section{Results and discussion}
\label{sec:results}
The accuracy of the FRPA method is evaluated by comparing to the ADC(3) method, and to Coupled Cluster calculations with Single, Double and Perturbative Triple [CCSD(T)] excitations. The latter method should be of a comparable level of theory as both the ADC(3) and FRPA. Where possible, the comparison with experimental results~\cite{CCCBDB} (or computational basis-set limits) is also made.
\subsection{Ground-State Energies and Ionization Energies at equilibrium geometry}
We first concentrate on calculating ground-state energies and ionization energies in equilibrium for a set of diatomic molecules with a singlet ground state. Calculations were performed for a number of different separation distances around the approximate equilibrium distance, after which a third-order polynomal was fitted to find the true energy minimum and equilibrium distance. The results calculated in a cc-pVDZ basis are presented in Table~\ref{tab:1a}.

\begin{table}[h]
\caption{FRPA results for some diatomic molecules and $\mathrm{BeH}_2$ in a cc-pVDZ basis set. The ground-state energy $E_0$ and vertical ionization energy $\mathrm{I}$ are in Hartree, equilibrium bond distance $r_0$ is in Angstrom. FRPA and FTDA refer to the calculations after the first iteration, while FRPAc and FTDAc refer to the calculations where consistency on the      Hartree-Fock level was applied. The calculated data are compared to the high-level ab-initio method CCSD(T) where available and to experimental data or exact calculations from Ref. \cite{CCCBDB}.}
\begin{tabular*}{\textwidth}{@{\extracolsep{\fill}} c  c  r  r  r  r  r  r  }
\hline\hline
Molecule &  & FTDA & FTDAc & FRPA & FRPAc & CCSD(T) & \multicolumn{1}{c}{Expt.}\\ 
\hline 

$\mathrm{H}_2$ & $E_0$ & $-1.170$ & $-1.161$ & $-1.170$ & $-1.161$ & $-1.164$ & $-1.175$\\
& $r_0$ & $0.769$ & $0.757$ & $0.770$ & $0.757$ & $0.761$ & $0.741$ \\
& I & $0.594$ & $0.589$ & $0.594$ & $0.589$ & $0.583$ & $0.591$\\

$\mathrm{HF}$ & $E_0$ & $-100.175$ & $-100.224$ & $-100.173$ & $-100.228$ & $-100.228$ & -\\
& $r_0$ & $0.904$ & $0.916$ & $0.897$ & $0.913$ & $0.920$ & $0.917$ \\
& I & $0.577$ & $0.577$ & $0.572$  & $0.571$ & $0.628$ & $0.592$\\

$\mathrm{HCl}$ & $E_0$ & $-460.295$ & $-460.256$ & $-460.293$ & $-460.258$ & $-460.254$ & -\\
& $r_0$ & $1.314$ & $1.297$ & $1.314$ & $1.293$ & $1.290$ & $1.275$ \\
& I & $0.457$ & $0.450$ & $0.457$ & $0.450$ & $0.471$ & -\\

$\mathrm{BF}$ &$E_0$ & $-124.331$ & $-124.365$ & $-124.332$ & $-124.368$ & $-124.380$ & -\\
& $r_0$ & $1.285$ & $1.284$ & $1.305$ & $1.285$ & $1.295$ & $1.267$ \\
& I & $0.417$ & $0.395$ & $0.431$ & $0.402$ & $0.406$ & -\\

$\mathrm{BeH}_2$ & $E_0$ & $-15.855$ & $-15.831$ & $-15.856$ & $-15.832$ & $-15.835$ & -\\
& $r_0$ & $2.747$ & $2.674$ & $2.766$ & $2.674$ & $2.678$ & $2.680$ \\
& I & $0.437$ & $0.433$ & $0.435$ & $0.432$ & $0.446$\footnotemark & -\\

$\mathrm{N}_2$ & $E_0$ & - & $-109.258$ & - & $-109.272$ & $-109.276$ & -\\
& $r_0$ & - & $1.104$ & - & $1.106$ & $1.119$ & $1.098$ \\
& I & - & $0.565$ & - & $0.544$ & $0.602$\footnotemark[1] & $0.573$ \\

$\mathrm{CO}$ & $E_0$ & $-113.096$  & $-113.037$ & $-113.100$ & $-113.048$ & $-113.055$ & -\\
& $r_0$ & $1.140$ & $1.130$ & $1.133$ & $1.123$ & $1.145$ & $1.128$\\
& I & $0.529$ & $0.503$ & $0.523$ & $0.494$ & $0.550$\footnotemark[1] & $0.515$\\

\hline\hline
\end{tabular*}

\footnotetext[1]{Only up to CCD level}
   \label{tab:1a}
\end{table}

The ground-state energies for the molecules $\mathrm{H}_2$ to $\mathrm{BeH}_2$ show little difference (at most $4$ mH) between ADC(3) and FRPA. The differences for $\mathrm{N}_2$ and $\mathrm{CO}$, containing double-triple bonds, are somewhat larger, of the order of $10$ mH. The FRPAc ground-state energies tend to be close to the CCSD(T) results with a maximum deviation of $12$ mH in case of BF. In general, ADC(3) deviates more from CCSD(T).

The equilibrium bond distances show a larger spreading. The equilibrium bond distances for ADC(3) and FRPA have comparable deviations from the experimental values, and in the majority of cases are closer to the experimental value than the CCSD(T) results. The same conclusion can be made for the ionization energies, for which ADC(3) and FRPA outperform the coupled cluster results, when the experimental value is available.

One remarkable fact is the lack of an equilibrium distance (no energy minimum) for $\mathrm{N}_2$ in both the ADC(3) and FRPA calculations without incorporating self-consistency at the level of the Hartree-Fock-like diagram. This example stresses the importance of a consistent treatment of the static self-energy. The inclusion of self-consistency in the calculations tends to adjust the results toward experiment, where needed.

\begin{table}[h]
   \caption{Vertical ionization energies in Hartree calculated in the aug-cc-pVDZ basis set. The values between braces are calculated without the $1\sigma_u$-level of $\mathrm{N}_2$. Experimental values are from Ref.~\cite{Trofimov2005}.}
\begin{tabular*}{\textwidth}{@{\extracolsep{\fill}} l c c c c c c}
\hline\hline
& & \multicolumn{4}{c}{FRPA}  & Expt.\\
& & cc-pVDZ & aug-cc-pVDZ & cc-pVTZ & aug-cc-pVTZ & \\ 
Molecule & Level \\
\hline
$\mathrm{HF}$ \\
& 1$\pi$ &                 15.46 &  16.06 &  16.18 &  16.33 &  16.05\\
& 3$\sigma$ &              19.57 &  20.01 &  20.06 &  20.21 &  20.00\\
\hline
\hline
\\
& & \multicolumn{4}{c}{FRPAc}  & Expt.\\
& & cc-pVDZ & aug-cc-pVDZ & cc-pVTZ & aug-cc-pVTZ & \\

Molecule & Level \\
\hline
$\mathrm{HF}$ \\
& 1$\pi$ &                 15.53 &  16.34 &  16.17 &  16.42 &  16.05\\
& 3$\sigma$ &              19.54 &  20.24 &  20.00 &  20.27 &  20.00\\
\hline\hline
\end{tabular*}
   \label{tab:2}
\end{table}

In order to compare with earlier ADC(3) calculations, we calculated vertical ionization energies for three diatomic molecules with the settings used in Ref.~\cite{Trofimov2005}, i.e. at the experimental bond length and with the aug-cc-pVDZ basis set. The results are presented in Table~\ref{tab:2}. The present FTDAc results are in good agreement with the Dyson ADC(3) results in Ref.~\cite{Trofimov2005}. The differences are less than $2$ mH and should probably be ascribed to a slightly different treatment of the HF-like self-energy. Compared to experiment, the mean absolute error is of the same order of magnitude for ADC(3) and FRPA. Note that there is a large deviation for the $2\sigma_u$-level of $\mathrm{N}_2$ in the FRPA which has a substantial influence on the mean error value. Apart from this level the mean absolute error of FTDAc and FRPAc is the same. 

\begin{table}
   \caption{Ground state energies and vertical ionization energies in Hartree for $\mathrm{HF}$, calculated in different basis sets. Experimental values are from Ref.~\cite{Trofimov2005}, CCSD(T) values are from Ref.~\cite{CCCBDB}.}
   \label{tab:3}
\begin{tabular*}{\textwidth}{@{\extracolsep{\fill}} l c c c c c c c}
\hline\hline
& & HF & FTDA & FTDAc & FRPA & FRPAc & Expt.\\
Molecule & Level & & & & & \\
\hline
$\mathrm{HF}$ & & & & & &\\
& 1$\pi$ &                 0.651 &  0.596 &  0.605 &  0.590 &  0.601 &  0.592\\
& 3$\sigma$ &              0.771 &  0.740 &  0.747 &  0.736 &  0.744 &  0.735\\
$\mathrm{CO}$ & & & & & & \\
& 5$\sigma$ &              0.555 &  0.532 &  0.510 &  0.528 &  0.503 &  0.515\\
& 1$\pi$ &                 0.641 &  0.626 &  0.622 &  0.623 &  0.619 &  0.621\\
& 4$\sigma$ &              0.808 &  0.737 &  0.739 &  0.715 &  0.720 &  0.724\\
$\mathrm{N}_2$ & & & & & &\\
& 3$\sigma_g$ &            0.634 &  0.593 &  0.575 &  0.579 &  0.558 &  0.573\\
& 1$\pi_u$ &               0.615 &  0.632 &  0.618 &  0.651 &  0.630 &  0.624\\
& 2$\sigma_u$ &            0.781 &  0.711 &  0.698 &  0.672 &  0.658 &  0.690\\
\\
& $\bar{\Delta}_{abs}$ (mH) &   49 (44) &  12 (10)  &  8 (8) & 10 (9) &  11 (8)\\
& $\Delta_{max}$ (mH)&        91 (84) &  21 (20) &  15 (15) &  27 (27) &  32 (15)\\
\hline\hline
\end{tabular*}
\end{table}

We also checked the basis-set dependency of the results in Tables~\ref{tab:1a}-\ref{tab:2} by performing calculations for HF in the cc-pVDZ, cc-pVTZ, aug-cc-pVDZ and aug-cc-pVTZ basis sets. The differences in ionization energies between DZ and TZ in Table~\ref{tab:3} are of the order of $25$ mH for the non-augmented and $10$ mH for the augmented basis sets. The convergence behavior of the ground-state energies calculated with FRPAc are very comparable to CCSD(T). The weaker convergence in FRPA again demonstrates the importance of self-consistency for the Hartree-Fock-like diagram.

\subsection{Dissociation problems for $\mathrm{H}_2$}
The FRPA fails to describe the correct dissociation behavior of diatomic molecules due to the appearance of instabilities in the RPA. The HF ground state becomes unstable with respect to ph-excitations in the dissociation limit. The RPA hamiltonian matrix is no longer positive-definite which results in complex solutions to the RPA equations. This is easily seen by analyzing $\mathrm{H}_2$ in a minimal basis set. The spatial wave functions are $1\mathrm{s}$ functions centered on the $\mathrm{H}$-atoms A and B. These can be put in a bonding and anti-bonding combination, which will be the Hartree-Fock hole and particle state
\begin{eqnarray}
|b) &=& \tfrac{1}{\sqrt{2}}\left[|A) + |B)\right]\nonumber\\
|a) &=& \tfrac{1}{\sqrt{2}}\left[|A) - |B)\right].
\end{eqnarray}
These states are normalized to unity at great separation, which is the case we are interested in.

The Hartree-Fock ground state is always the spatially symmetric state with positive parity and 0 spin:
\begin{eqnarray}
|\Phi_{0}^{HF}> &=& |bb)\tfrac{1}{\sqrt{2}}\left[|\uparrow\downarrow)-|\downarrow\uparrow)\right]\\
&=& a^{\dagger}_{b\uparrow}a^{\dagger}_{ b\downarrow}|0>.
\end{eqnarray}
The possible ph-excitations can only be formed by removing a bonding state and replacing it with an anti-bonding state. This results in a spin singlet and triplet. The energy of the triplet state 
\begin{equation}
|\Phi> = \left[ a^{\dagger}_{a} \otimes a^{\dagger}_{b}\right]_{M_S}^{S=1} |0>
\end{equation}
is found to be
\begin{equation}
<\Phi|H|\Phi> = (ab|H|ab)-(ab|H|ba).
\end{equation}
In the dissociation limit, the overlap and interaction matrix elements between the $1\mathrm{s}$ wavefunctions for hydrogens A and B vanish, and the energy of the triplet state simply becomes
\begin{equation}
<\Phi|H|\Phi> = (AB|H|AB) + (BA|H|BA).
\end{equation}
This is exactly the energy one would expect for the dissociation state where each hydrogen atom receives one electron, which is the exact ground state when the two hydrogens are separated by a large distance. The energy of this state is thus the exact ground-state energy, and automatically lower than the Hartree-Fock ground-state energy in this limit. As a result, a negative phonon energy occurs for this triplet state in the ph-TDA. In ph-RPA, the same mechanism gives rise to a complex phonon energy.

\begin{figure}
   \begin{center}
      \subfigure[]{
         \includegraphics[height=0.48\textwidth,angle=270]{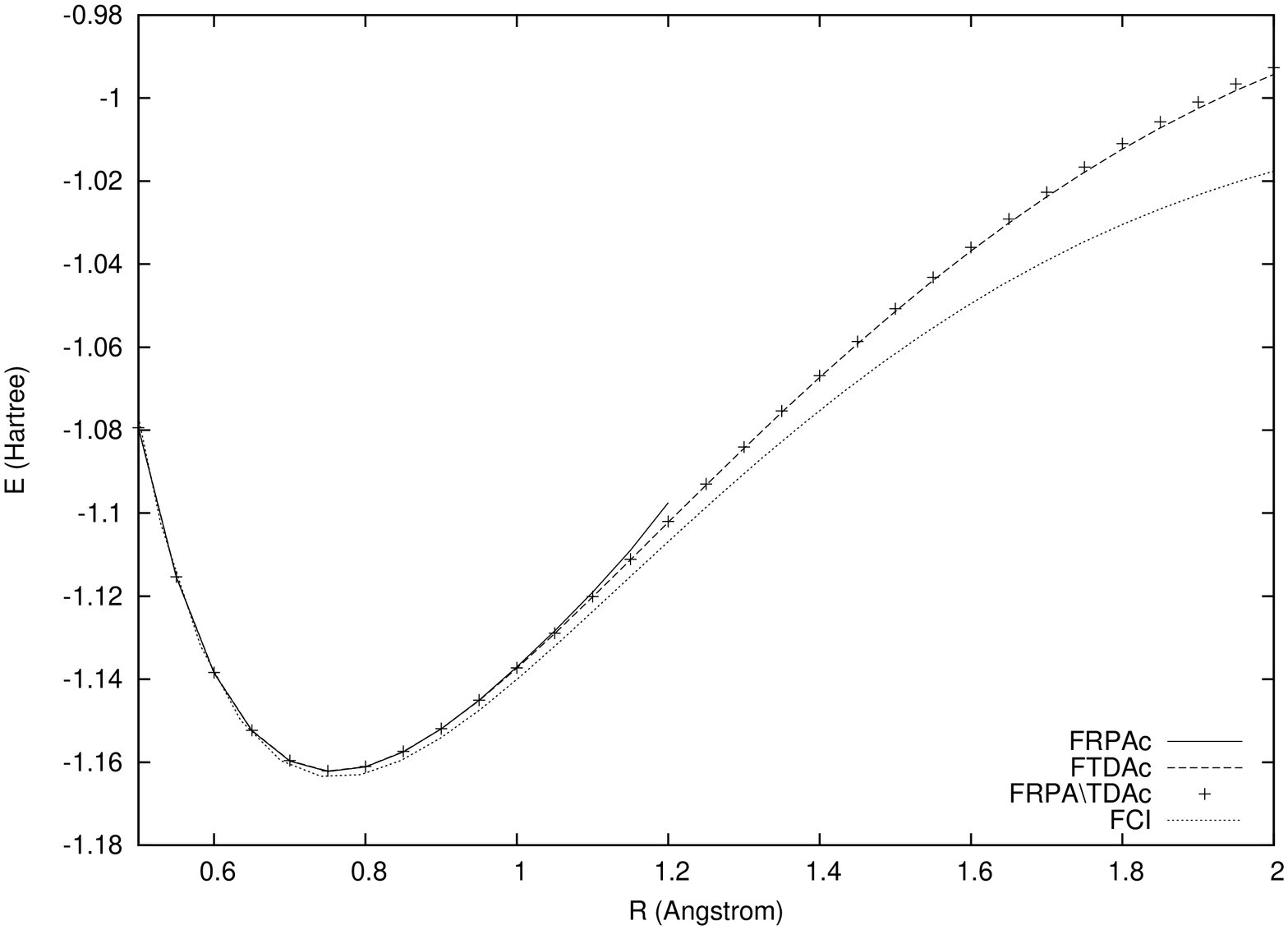}
         \label{fig:instabE0}
      }
      \subfigure[]{
         \includegraphics[height=0.48\textwidth,angle=270]{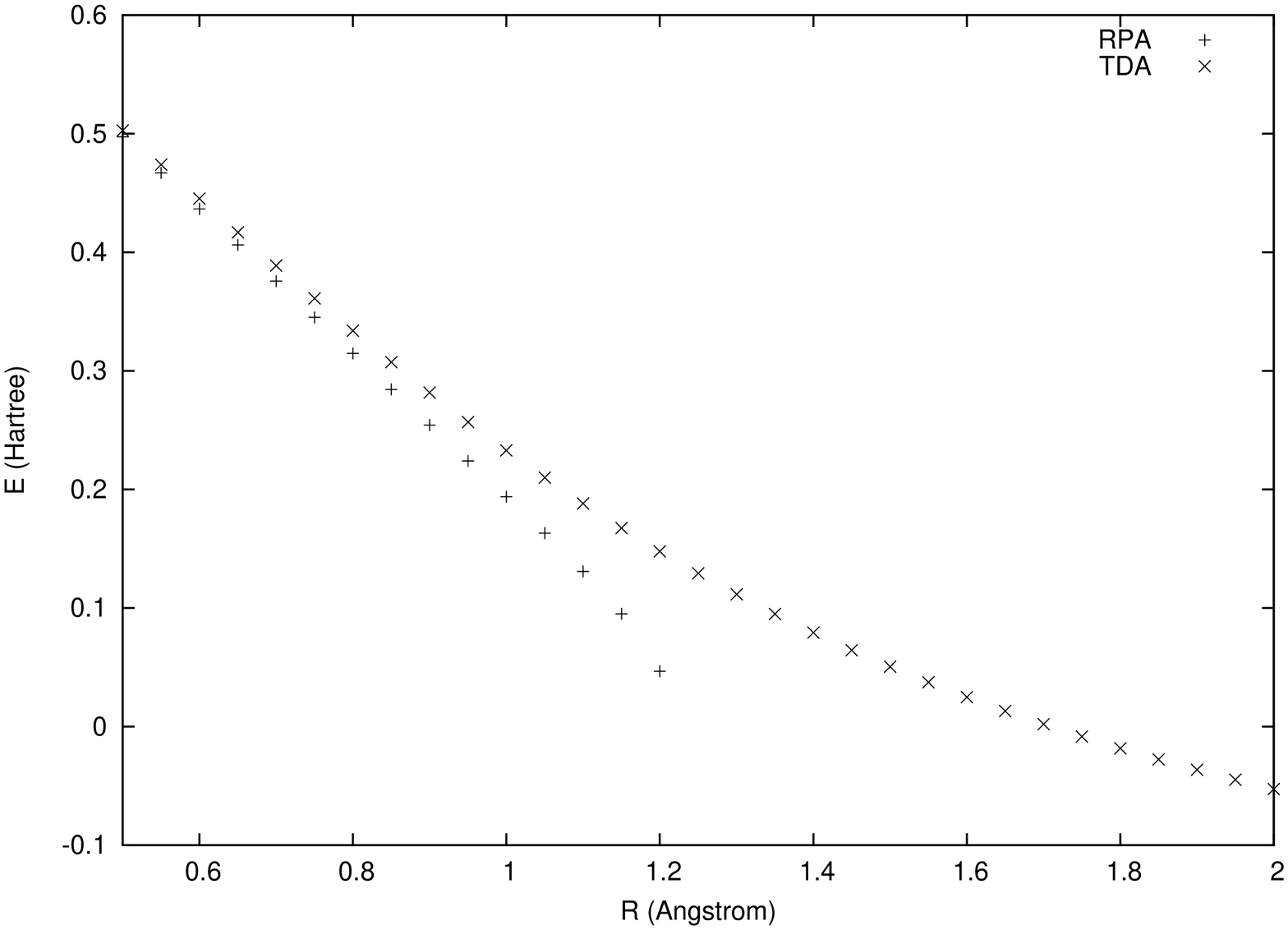}
         \label{fig:instabEx}
      }
      \caption{Demonstration of the problems in the dissociation limit for $\mathrm{H}_2$ in the cc-pVDZ basis set. \subref{fig:instabE0} shows the ground-state energy for $\mathrm{H}_2$ calculated with the FRPA (solid line), FTDA (dashed line) and the mixed procedure (crosses) where the spin-1 channel of the RPA phonons has been replaced with TDA phonons compared to the exact full-CI result (dotted line). \subref{fig:instabEx} shows the lowest ph-RPA and ph-TDA excitation energy in the spin-1 channel for $\mathrm{H}_2$ as a function of internuclear distance.}
      \label{fig:instab}
   \end{center}
\end{figure}

This behavior is actually found both in the minimal basis set model for $\mathrm{H}_2$ as in more realistic calculations. As an example of this behavior we have plotted the ground state energy for $\mathrm{H}_2$ calculated in the cc-pVDZ basis set in Figure~\ref{fig:instabE0}. For distances larger than approximately $1.2\text{ \AA}$ ph-RPA becomes unstable. At this distance the lowest ph-RPA eigenvalue in the spin-1 channel becomes zero, as can be seen in Figure~\ref{fig:instabEx}. Beyond this distance ph-RPA acquires a complex eigenvalue. The ph-TDA eigenvalue becomes negative as well, which is unphysical for an excitation energy but does not pose any computational problems. A possible solution is to substitute the problematic spin-1 ph-RPA channel with its ph-TDA counterpart. The procedure then remains stable but is almost identical to FTDA. In any case, the use of TDA phonons does not guarantee a correct dissociation limit. Both the mixed and the pure TDA Faddeev method deviate substantially from the exact full-configuration-interaction results in the dissociation limit. A more fundamental solution to this problem would probably be the fully self-consistent approach, where the propagator is allowed to have fragmented spectral strength~\cite{Barbieri2003}. This, however, implies a huge computational effort which lies beyond the scope of this paper.\\

\section{Conclusion}
\label{sec:conclusion}
In this work we have investigated the application of the FRPA technique to small molecules. The computational cost of this method is not much higher than that of the more established ADC(3) method and in any case lower than the cost for CCSD(T). The results at equilibrium geometry are comparable in accuracy to the ones obtained with the ADC(3) method which is in line with the earlier atomic calculations.

The self-consistent treatment of the Hartree-Fock diagram has a positive effect on the numerical results and should always be included. The possibility of complex eigenvalues in the RPA and FRPA eigenvalue equations is a problem that has to be kept in mind. We have shown that RPA instabilities are bound to occur in the dissociation limit, when the Hartree-Fock propagator is used as a starting point. A possible way out is to increase the self-consistency by allowing propagators with fragmented single-particle strength, which will be the object of future research.

\begin{acknowledgments}
M.D. is supported by a Ph.D. grant provided by FWO-Flanders (Fund for Scientific Research). C.B. acknowledges the Japanese Ministry of Education, Science and Technology (MEXT) under KAKENHI grant no. 21740213.
\end{acknowledgments}

\appendix

\section{Derivation of the FRPA equations for i=3}
\label{app:frpa}

The product of the forward propagating uncorrelated 2p1h propagator and the interaction vertex is needed to find an expression in function of RPA-amplitudes and the two-particle interaction. We will do this for the case $i=3$, the other two cases are equivalent, but involving the $\Gamma^{ph}$ instead of the $\Gamma^{pp}$. The combination of the free 2p1h propagator and the vertex function can be written as
\begin{eqnarray}
\left[G^{(0)>}\Gamma^{(3)}\right)_{\alpha\beta\gamma,\lambda\mu\nu}\left(E\right] &=& \frac{1}{2}\int \frac{\mathrm{d}E_1}{2\pi i}\int \frac{\mathrm{d} E_2}{2\pi i} \sum_{\rho\sigma} G^{(0)>}_{\alpha, \rho}\left(E_2\right)G^{(0)>}_{\beta,\sigma}\left(E_1-E_2\right)\nonumber\\
&&\quad G^{(0)<}_{\gamma,\nu}\left(E_1-E\right)\Gamma^{pp}_{\rho\sigma,\lambda\mu}\left(E_1\right)\\
&=& \frac{\delta_{\gamma\lambda}}{2}\int \frac{\mathrm{d}E_1}{2\pi i}\frac{1}{E_1-E-\epsilon_{\gamma}-i\eta} \Gamma^{pp}_{\alpha\beta,\lambda\mu}\left(E_1\right)\nonumber\\
&&\quad\int \frac{\mathrm{d} E_2}{2\pi i} \frac{1}{E_2-\epsilon_{\alpha}+i\eta}\frac{1}{E_1 - E_2-\epsilon_{\beta}+i\eta}\nonumber\\
&=& \frac{\delta_{\gamma\lambda}}{2}\int \frac{\mathrm{d}E_1}{2\pi i}\frac{1}{E_1-E-\epsilon_{\gamma}-i\eta} \Gamma^{pp}_{\alpha\beta,\lambda\mu}\left(E_1\right)\frac{1}{E_1-\epsilon_{\alpha}-\epsilon_{\beta} +i\eta}\nonumber.
\end{eqnarray}
Here the explicit expression for the phonon propagator is needed.
\begin{eqnarray}
\Gamma^{pp}_{\alpha\beta,\lambda\mu}\left(E\right)  &=& V_{\alpha\beta,\lambda\mu} + \sum_{\rho,\sigma,\xi,\chi}V_{\alpha\beta,\rho\sigma}G^{pp}_{\rho\sigma,\xi\chi}\left(E\right)V_{\xi\chi,\lambda\mu}\\
&=& V_{\alpha\beta,\lambda\mu} + \sum_{\rho,\sigma,\xi,\chi}V_{\alpha\beta,\rho\sigma}\left(\sum_m \frac{\mathcal{X}_{\rho\sigma,m}^{pp}\mathcal{X}_{\xi\chi,m}^{pp \dagger}}{E-\epsilon^{pp+}_m+i\eta} - \sum_n \frac{ \mathcal{Y}^{pp}_{\xi\chi,n}\mathcal{Y}^{pp\dagger}_{\rho\sigma,n}}{ E-\epsilon^{pp-}_n-i\eta}\right) V_{\xi\chi,\lambda\mu}\nonumber\\
&=& V_{\alpha\beta,\lambda\mu} + \sum_m \frac{\Delta_{\alpha\beta,m}^{pp+}\Delta_{\lambda\mu,m}^{pp+ \dagger}}{E-\epsilon^{pp+}_m+i\eta} - \sum_n \frac{ \Delta^{pp+}_{\lambda\mu,n}\Delta^{pp+\dagger}_{\alpha\beta,n}}{E-\epsilon^{pp-}_n-i\eta}\nonumber
\end{eqnarray} 
The $\Delta^{pp}$ are introduced as the product between the interaction and the normal RPA-amplitudes $\mathcal{X}^{pp}$ and $\mathcal{Y}^{pp}$. Due to the RPA-equations~(\ref{eqn:RPApp}), this correspondence can also be expressed as
\begin{eqnarray}
\Delta^{pp+}_{\alpha\beta,m} &=& \frac{\mathcal{X}^{pp}_{\alpha\beta,m}}{\sqrt{2}\left(\epsilon_m^{pp+}-\epsilon_{\alpha}-\epsilon_{\beta}\right)}\\
\Delta^{pp-}_{\alpha\beta,n} &=& \frac{\mathcal{Y}^{pp}_{\alpha\beta,m}}{\sqrt{2}\left(\epsilon_n^{pp-}-\epsilon_{\alpha}-\epsilon_{\beta}\right)}\nonumber,
\end{eqnarray}
where the factor $\frac{1}{\sqrt{2}}$ arises from the normalization condition for the pp RPA amplitudes and is not needed in case of ph RPA.\\
After performing the necessary integrations over the intermediate energies, one arrives at
\begin{eqnarray}
\left[G^{(0)>}\Gamma^{(3)}\right]_{\alpha\beta\gamma,\lambda\mu\nu}\left(E\right) &=& \frac{1}{2}\frac{\delta_{\gamma\nu}}{E-\epsilon_{\alpha}-\epsilon_{\beta} + i \eta}\left(V_{\alpha\beta,\lambda\mu} + \sum_n \frac{\Delta^{pp+\dagger}_{\alpha\beta,n} \Delta^{pp-}_{\lambda\mu,n}}{E-\left(\epsilon_n^{pp+}-\epsilon_{\gamma}\right) + i\eta} + \right.\nonumber\\
&&\quad \left.\sum_m \frac{\Delta^{pp-}_{\alpha\beta,m}\Delta^{pp-\dagger}_{\lambda\mu,m}\left(E-\epsilon_{\alpha} - \epsilon_{\beta} + \epsilon_{\gamma} -\epsilon_{\lambda}-\epsilon_{\mu} + \epsilon^{pp-}_{m} \right)}{\left(\epsilon_m^{pp-}-\epsilon_{\alpha}-\epsilon_{\beta}\right)\left(\epsilon_m^ {pp-} -\epsilon_{\lambda}-\epsilon_{\mu}\right)}\right)\\
&=& \frac{\delta_{\gamma\nu}}{2}\left(\sum_n \frac{\Delta^{pp+\dagger}_{\alpha\beta,n}\Delta^{pp+}_{\lambda\mu,n}}{\left(\epsilon_n^{pp+}-\epsilon_{\alpha}-\epsilon_{\beta}\right)\left(E-\epsilon_n^{pp+}+\epsilon_{\gamma}\right)}\right.\nonumber\\
&&\quad \left. + \sum_m \frac{\Delta^{pp-}_{\alpha\beta,m}\Delta^ {pp-\dagger}_{\lambda\mu,m}}{\left(\epsilon_m^{pp-}-\epsilon_{\alpha}-\epsilon_{\beta}\right)\left(\epsilon_{m}^{pp-}-\epsilon_{\lambda}-\epsilon_{\mu}\right)} \right)\nonumber\\
&=& \frac{1}{2}\delta_{\gamma\nu}\left(\sum_n \mathcal{X}^{pp+\dagger}_{\alpha\beta,n}\frac{1}{E-\epsilon_{n}^{pp+}+\epsilon_{\gamma}}\mathcal{X}^{pp+}_{\lambda\mu,n}\left(\epsilon_{n}^{pp+}-\epsilon_{\lambda}-\epsilon_{\mu}\right)\right. \nonumber\\
&& \quad \left. + \sum_m \mathcal{Y}^{pp-}_{\alpha\beta,m}\mathcal{Y}^{pp-\dagger}_{\lambda\mu,m}\right)\nonumber,
\end{eqnarray}
where in the second transition the property $\Gamma^{pp}_{\alpha\beta,\lambda\mu}\left(\epsilon_{\alpha} + \epsilon_{\beta}\right) = 0$ was used to simplify the relation.\\

\section{ADC(3) as special case of FRPA}
\label{app:adc}

To show that ADC(3) is incorporated in FRPA one has to change the RPA-interactions with TDA-interactions. This can be done by setting the off-diagonal blocks in equations~(\ref{eqn:RPApp}) and~(\ref{eqn:RPAph}) to zero. As a result there are no backward propagating amplitudes $\mathcal{Y}$. The FRPA-equation~(\ref{eqn:FRPAham}) simplifies due to the disappearance of the $HH^{\dagger}$. After projecting out the spurious solutions we get the equation
\begin{equation}
\epsilon^{Fd}\mathcal{X}= \frac{1}{6}\left(\dblone-\dblone_{ex}\right)\left(\sum_{i=1,2,3} U^{(i)}D^{(i)}U^{(i)-1} + 2U^{(i)}T^{(i)\dagger}\right)\left(\dblone-\dblone_{ex}\right)\mathcal{X}.
\end{equation}
As an example we will again work out the term for $i=3$ for the 2p1h energies
\begin{eqnarray}
\left(U^{(3)}D^{(3)}U^{(3)-1}+2U^{(3)}T^{(3)\dagger}\right)_{\alpha\beta\gamma,\lambda\mu\nu} &=& \delta_{\gamma\nu}\sum_{n}\mathcal{X}^{pp+}_{\alpha\beta,n}\left(\epsilon_n^{pp+}-\epsilon_{\mu}\right)\left(\mathcal{X}^{pp+}\right)^{-1}_{\lambda\mu,n} \nonumber\\
&&\quad+ 2\delta_{\gamma\nu}\sum_{n}\mathcal{X}_{\alpha\beta,n}^{pp+}\left(\epsilon_n^{pp+}-\epsilon_{\alpha}-\epsilon_{\beta}\right)\mathcal{X}^{pp+}_{\lambda\mu,n}.
\end{eqnarray}
By eliminating the TDA eigenvalues using their generating equations
\begin{equation}
\epsilon_n^{pp+}\mathcal{X}_{\alpha\beta,n}^{pp+} = \left(\epsilon_{\alpha}+\epsilon_{\beta}\right)\left(\mathcal{X}_{\alpha\beta,n}^{pp+}-\mathcal{X}_{\beta\alpha,n}^{pp+}\right)+\frac{1}{2}\sum_{\lambda,\mu}V_{\alpha\beta,\lambda\mu}\mathcal{X}_{\lambda\mu,n}^{pp+}
\end{equation}
and using the orthonormality of the TDA eigenvectors
\begin{equation}
\sum_n\mathcal{X}_{\alpha\beta,n}^{pp+}\mathcal{X}_{\lambda\mu,n}^{pp+} = \frac{\left(\delta_{\alpha\lambda}\delta_{\beta\mu}-\delta_{\alpha\mu}\delta_{\beta\lambda}\right)}{2},
\end{equation}
we arrive at
\begin{equation}
\left(U^{(3)}D^{(3)}U^{(3)-1}+                                                 2U^{(3)}T^{(3)\dagger}\right)_{\alpha\beta\gamma,\lambda\mu\nu} = \frac{\delta_{\gamma\nu}}{2}\left[\left(\delta_{\alpha\lambda}\delta_{\beta\mu}-\delta_{\alpha\nu}\delta_{\beta\lambda}\right)\left(\epsilon_{\alpha}+\epsilon_{\beta}-\epsilon_{\gamma}\right) + 3V_{\alpha\beta,\lambda\mu}\right].
\end{equation}
Similar steps have to be taken for the other two channels. The sum of the three channels after anti-symmetrization becomes
\begin{align}
&\left[\frac{1}{6}\left(\dblone-\dblone_{ex}\right)\left(\sum_{i=1,2,3}U^{(i)}D^{(i)}U^{(i)-1}+                                                                2U^{(i)}T^{(i)\dagger}\right)\left(\dblone-\dblone_{ex}\right)\right]_{\alpha\beta\gamma,\lambda\mu\nu}  \hfill \nonumber\\
&=\delta_{\gamma\nu}\left(\delta_{\alpha\lambda}\delta_{\beta\mu}-\delta_{\alpha\mu}\delta_{\beta\lambda}\right)\left(\epsilon_{\alpha}+\epsilon_{\beta}-\epsilon_{\gamma}\right)\nonumber\\
&\quad+ \delta_{\gamma\nu}V_{\alpha\beta,\lambda\mu}+\delta_{\alpha\lambda}V_{\beta\nu,\mu\gamma} + \delta_{\beta\mu}V_{\alpha\nu,\lambda\gamma} - \delta_{\alpha\mu}V_{\beta\nu,\lambda\gamma} - \delta_{\beta\lambda}V_{\alpha\nu,\mu\gamma}.
\label{eqn:adc3}
\end{align}
This is exactly the same expression as in ADC(3). The Faddeev Tamm Dancoff Approximation (FTDA) and ADC(3) are completely equivalent.

\end{document}